\documentclass[prm,twocolumn,showpacs,preprintnumbers,amsmath,amssymb,superscriptaddress,longbibliography]{revtex4-1}
\usepackage{dcolumn}
\usepackage{bm,graphicx}
\usepackage{etoolbox}
\apptocmd{\thebibliography}{\raggedright}{}{}
\usepackage{url}
\usepackage[colorlinks]{hyperref}
\usepackage{breakurl}
\usepackage{color}
\usepackage{soul}
\begin{document}

\title{3D Dirac semimetal $\mbox{Cd}_3 \mbox{As}_2$: a review of material properties}

\author{I.~Crassee}
\affiliation{Laboratoire National des Champs Magn\'etiques Intenses, CNRS-UGA-UPS-INSA, 25, avenue des Martyrs, 38042 Grenoble, France}

\author{R.~Sankar}
\affiliation{Institute of Physics, Academia Sinica, Nankang, 11529 Taipei, Taiwan}

\author{W.-L.~Lee}
\affiliation{Institute of Physics, Academia Sinica, Nankang, 11529 Taipei, Taiwan}

\author{A.~Akrap}
\affiliation{University of Fribourg, Department of Physics, Chemin du Mus\'ee 3, CH-1700 Fribourg, Switzerland}

\author{M.~Orlita}\email{milan.orlita@lncmi.cnrs.fr}
\affiliation{Laboratoire National des Champs Magn\'etiques Intenses, CNRS-UGA-UPS-INSA, 25, avenue des Martyrs, 38042 Grenoble, France}
\affiliation{Institute of Physics, Charles University, Ke Karlovu 5, 12116 Praha 2, Czech Republic}

\date{\today}

\begin{abstract} Cadmium arsenide (Cd$_3$As$_2$) -- a time-honored and widely explored material in solid-state physics -- has recently attracted considerable attention. This was triggered by a theoretical prediction concerning the presence of 3D symmetry-protected massless Dirac electrons, which could turn Cd$_3$As$_2$ into a 3D analogue of graphene.
Subsequent extended experimental studies have provided us with compelling experimental evidence of conical bands in this system, and revealed a number of interesting properties and phenomena. At the same time, some of the material properties remain the subject of vast discussions despite recent intensive experimental and theoretical efforts, which may hinder the progress in understanding and applications of this appealing material. In this review, we focus on the basic material parameters and properties of Cd$_3$As$_2$, in particular those which are directly related to the conical features in the electronic band structure of this material. The outcome of experimental investigations, performed on Cd$_3$As$_2$ using various spectroscopic and transport techniques within the past sixty years, is compared with theoretical studies. These theoretical works gave us not only simplified effective models, but more recently, also the electronic band structure calculated numerically using ab initio methods.
\end{abstract}

\maketitle

\section{Introduction}

Cadmium arsenide (Cd$_3$As$_2$) is an old material for condensed-matter physics, with its very first investigations dating back to the thirties~\cite{StackelbergZP35}.
Research on this material then continued extensively into the sixties and seventies, as reviewed in Ref.~\cite{ZdanowiczARMS75}. This is when the physics of semiconductors, those with a sizeable, narrow, but also vanishing energy band gap, strongly developed.

\begin{figure}
	\includegraphics[width=0.38\textwidth,trim={0cm 0cm 0cm 0cm}]{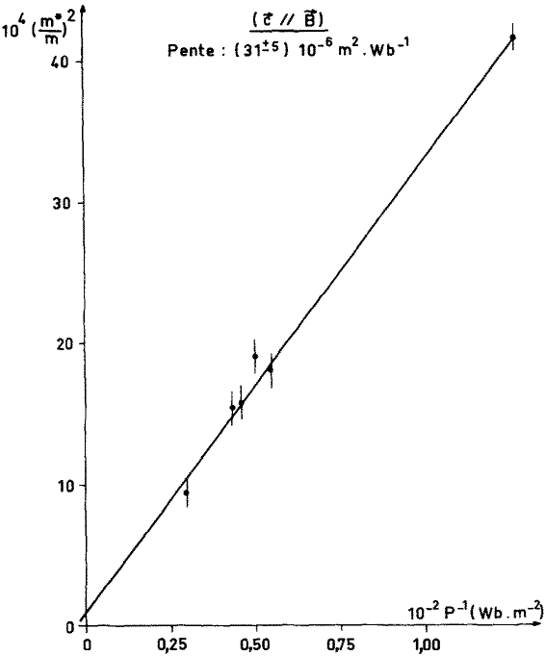}
	\caption{One of the very first experimental indications of a conical band in Cd$_3$As$_2$ presented by Rosenman~\cite{RosenmanJPCS69}: the square of the cyclotron mass $m^*$,
determined from the thermal damping of Shubnikov-de Haas oscillations, as a function of their inverse oscillation period $1/P$. The theoretically expected dependence for a conical band, $m^2=2e\hbar/(Pv^2)$, allows us to estimate the velocity parameter $v=(1.1\pm0.1)\times10^6$~m/s from the slope indicated in the plot. Reprinted  from~\cite{RosenmanJPCS69}
with permission by Elsevier, copyright (1969). \label{Rosenman}}
\end{figure}

In the early stages of research on Cd$_3$As$_2$, it was the extraordinarily high mobility of electrons, largely exceeding 10$^{4}$~cm$^{2}$/(V.s) at room temperature~\cite{RosenbergJAP59,TurnerPR61,RosenmanJPCS69},  that already attracted significant attention to this material. Even today, with a declared mobility well above 10$^{6}$~cm$^{2}$/(V.s)~\cite{LiangNatureMater14} at low temperatures, cadmium arsenide belongs to the class of systems with the highest electronic mobilities, joining materials such as graphene, graphite, bismuth and 2D electron gases in GaAs/GaAlAs heterostructures~\cite{MichenaudJPC72,Brandt88,HwangPRB08,BolotinPRL08,NeugebauerPRL09}.
Another interesting -- and in view of recent developments, crucial -- observation was the strong dependence of
the effective mass of electrons on their concentration, which implies a nearly conical conduction band, in other words the energy dispersion is linear in momentum, see Refs.~\cite{ArmitagePLA68,RosenmanJPCS69,RogersJPD71} and Fig.~\ref{Rosenman}.

To understand the electronic properties of Cd$_3$As$_2$, and its extraordinarily large mobility of electrons in particular, simple theoretical models for the
electronic band structure have been proposed in the standard framework of semiconductors physics, and compared with optical and transport experiments. In the initial phase of research, cadmium arsenide was treated as an ordinary Kane-like semiconductor or semimetal \cite{CaronPRB77,Bodnar77}. It was seen as a material with an electronic band structure that
closely resembled zinc-blende semiconductors with a relatively narrow or vanishing band gap.
However, no clear consensus was achieved concerning the particular band structure parameters, Refs.~\cite{ZdanowiczARMS75} or \cite{Blom80}.
Most notably, there was disagreement about the ordering of electronic bands and the presence of another conduction band at higher energies.

Renewed interest in the electronic properties of Cd$_3$As$_2$ was provoked by a theoretical study where Wang~\emph{et al.}~\cite{WangPRB13} invoked the presence of a pair of symmetry-protected 3D Dirac cones. This way, Cd$_3$As$_2$ would fit into an emergent class of materials which are nowadays referred to as
3D Dirac semimetals, for review see, \emph{e.g.}, Ref.~\cite{ArmitageRMP18}. These systems host 3D massless electrons described by the Dirac equation for particles
with a vanishing rest mass, thus implying a conical band twice degenerated due to spin. The possible lack of inversion symmetry may lift this degeneracy and transform Cd$_3$As$_2$ into a 3D Weyl semimetal, with two pairs of Weyl cones. The lack or presence of inversion symmetry in Cd$_3$As$_2$ is still not resolved~\cite{SteigmannACB68,AliIC14,DesratPRB18}.

\begin{figure}
	\includegraphics[width=0.43\textwidth,trim={0cm 0cm 0cm 0cm}]{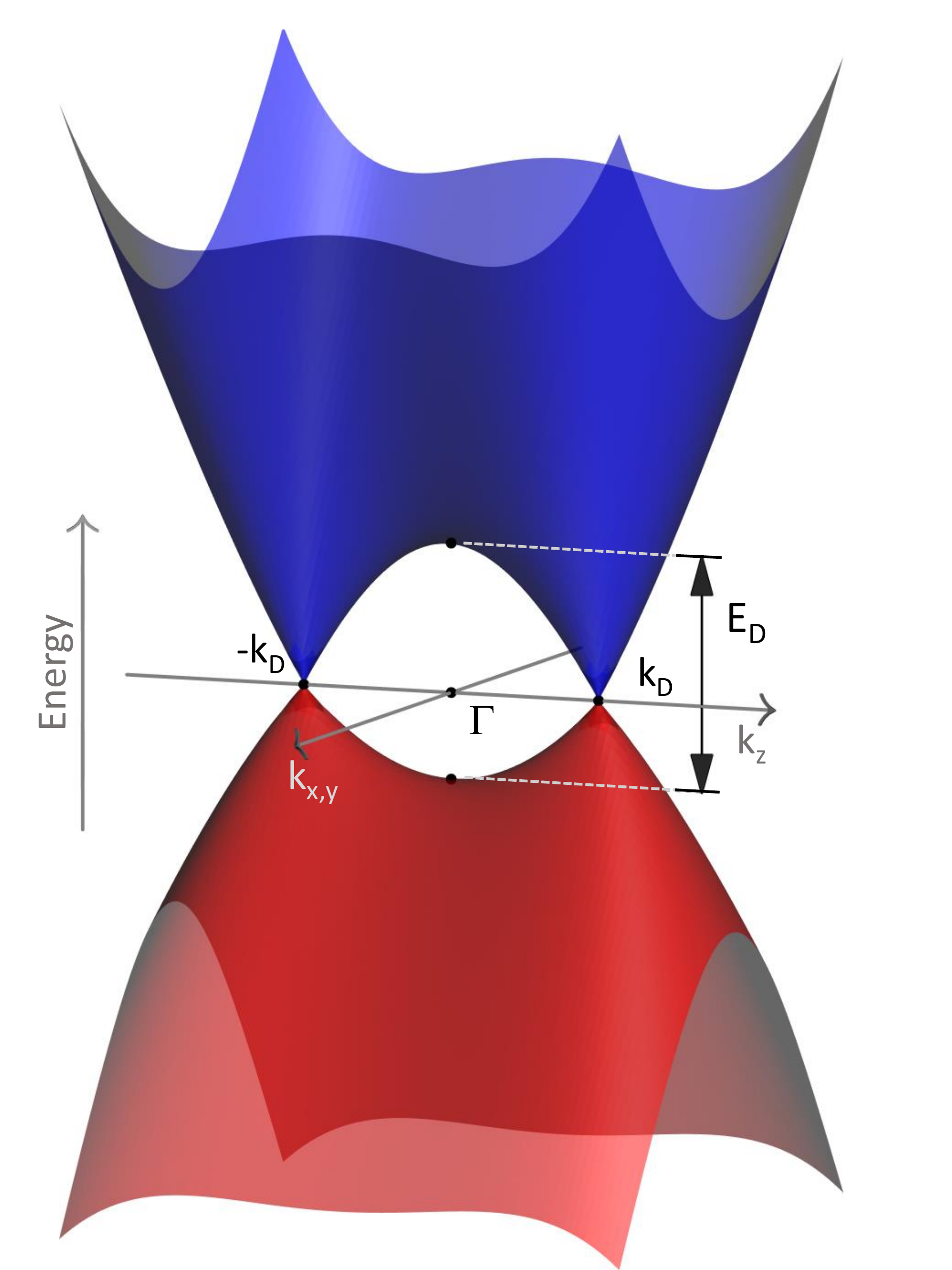}
	\caption{A schematic view of the 3D Dirac cones in Cd$_3$As$_2$. Two symmetry-protected cones emerge
in the vicinity of the $\Gamma$ point with the apexes located at the momenta $k=\pm k_D$ along the tetragonal axis of Cd$_3$As$_2$ (often associated with the $z$-axis).
In general, the Dirac conical bands are characterized by an anisotropic velocity parameter ($v_x=v_y\neq v_z$) and possibly also
by an electron-hole asymmetry. The energy scale of these cones (its upper bound) is given
by the $E_D$ parameter, which refers to the energy distance between the upper and lower Lifshitz points. \label{Scheme}}
\end{figure}

The theoretical prediction by Wang \emph{et al.}~\cite{WangPRB13} can be illustrated using a simple, currently widely accepted cartoon-like picture depicted in Fig.~\ref{Scheme}. It shows two 3D Dirac nodes, the points where the conical conduction and valence bands meet, located at the tetragonal $z$ axis of the crystal. Their protection is ensured by $C_4$ rotational symmetry~\cite{YangNatureComm14}.
With either increasing or decreasing energy, these two Dirac cones approach each other and merge into a single electronic band centered around the $\Gamma$ point, passing through a saddle-like Lifshitz point.

The prediction of a 3D Dirac semimetallic phase in a material, which is not only well-known in solid-state physics but also relatively stable under ambient conditions,
stimulated a considerable experimental effort.
The first angular-resolved photoemission spectroscopy (ARPES) and scanning tunneling microscopy/spectroscopy (STM/STS) experiments~\cite{LiuNatureMater14,JeonNatureMater14,BorisenkoPRL14,NeupaneNatureComm14} confirmed the presence of widely extending conical features in the band structure, creating a large wave of interest.
This wave gave rise to a number of experimental and theoretical studies, complementing those performed in the past, forming a rich knowledge of this material. As a result, cadmium arsenide now is among the most explored materials in current solid-state physics. 

In addition to the extraordinarily high electronic mobility, cadmium arsenide exhibits a very strong linear magneto-resistance ~\cite{LiangNatureMater14}, an anomalous Nernst effect~\cite{LiangPRL17}, and quantum Hall effect signatures when thinned down~\cite{UchidaNC17,SchumannPRL18}.
Furthermore, indications of the chiral anomaly, planar Hall effect, and electron transport through surface states have been reported~\cite{LiNatureComm15,MollNature16,JiaNatureComm16,WuPRB18}. Currently, these effects are all considered to be at least indirectly related to the specific relativistic-like band structure of this material. Yet surprisingly, consensus about the complex electronic band structure of Cd$_3$As$_2$ has not yet been fully established. 

A detailed understanding of the basic material parameters of Cd$_3$As$_2$ became essential for the correct interpretation of a wide range of observed, and yet to be discovered, physical phenomena. In this paper we review the the current knowledge of Cd$_3$As$_2$. We start with the properties of the crystal lattice and continue with the technological aspects of Cd$_3$As$_2$ growth. This is followed by a discussion of the theoretical and experimental investigation results, providing different possible views of this material's electronic bands. In particular, we focus on Dirac-like nodes, the most relevant aspect of the band structure when taking into consideration recent developments in this material's field.

\section{Crystal lattice}

\begin{figure}
	\includegraphics[width=0.48\textwidth,trim={0cm 0cm 0cm 0cm}]{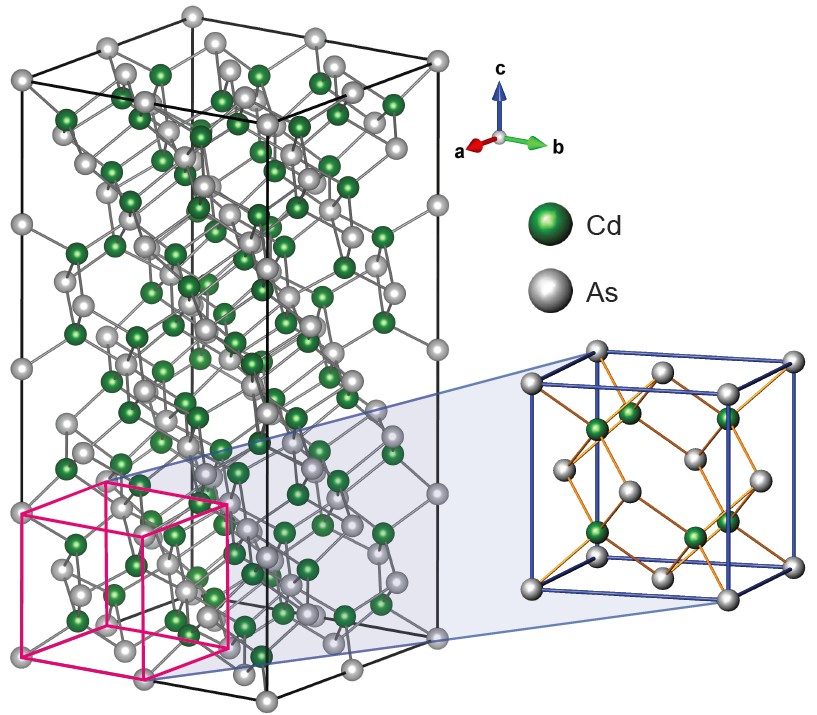}
	\caption{The non-primitive tetragonal unit cell of Cd$_3$As$_2$ is composed of $2\times2\times4$ weakly distorted antiflourite cells with two cadmium vacancies. This cell hosts
96 cadmium atoms and 64 arsenic atoms. \label{Lattice}}
\end{figure}

Cadmium arsenide has a relatively complex crystal structure, with 160 atoms in the unit cell. It has been the subject of several $x$-ray studies, resulting in somewhat contradicting conclusions~\cite{StackelbergZP35,SteigmannACB68,PietraszkoPSS73,AliIC14}. The current consensus implies that, at room and lower temperatures, which are the most relevant ones for fundamental and applied research conducted on this material, Cd$_3$As$_2$ is a tetragonal material with a unit cell of $a = b \approx 1.26$~nm and $c \approx 2.54$~nm. The particular space group remains a subject of discussion. More recent investigations favor a centrosymmetric group I4$_1$/acd (No. 142) ~\cite{PietraszkoPSS73,AliIC14} over a non-centrosymmetric space group  I4$_1$cd (No. 110), as previously suggested~\cite{SteigmannACB68}. The correct space group assignment is of considerable importance for the complete understanding of Cd$_3$As$_2$. When space inversion symmetry is not present, spin degeneracy is lifted, and the Dirac nodes possibly split into pairs of Weyl nodes. 

It is also important to note that the Cd$_3$As$_2$ crystal lattice, though clearly tetragonal, remains nearly cubic ($2a=2b\approx c$)~\cite{AliIC14}. The lattice may therefore be seen, in the simplest approach, as being composed of antifluorite (cubic) cells with two missing cadmium cations (Fig.~\ref{Lattice}). Due to the cadmium vacancy ordering, a very large unit cell of Cd$_3$As$_2$ is formed. Composed of $2\times2\times 4$ of antifluorite cells, it is oriented along the tetragonal $c$-axis, and contains 160 atoms (96 cadmium and 64 arsenic). Additionally, each single anti-fluorite cell is tetragonally distorted, with a small elongation along the $c$-axis ($c/2a \approx 1.006$)~\cite{ArushanovPCGC80,AliIC14}. This simplified image of nearly cubic antifluourite cells, serving as building blocks for the entire Cd$_3$As$_2$ lattice, appears as a useful starting point for simplified effective models and {\em ab initio} calculations.  Both are briefly discussed below. 

Finally, let us mention that above room temperature the crystal lattice of Cd$_3$As$_2$ undergoes a sequence of polymorphic phase transitions. The corresponding space group remains tetragonal but changes to P4$_2$/nbc (No.~133) around 220$^\circ$C and to P4$_2$/nmc (No.~137) around 470$^\circ$C~\cite{PietraszkoPSS73,ArushanovPCGC80}. At temperatures above 600$^\circ$C, the symmetry of the crystal changes to a cubic one, characterized by the space group Fm$\bar{3}$m (No.~225)~\cite{AliIC14}. Each of these phase transitions is accompanied by an abrupt change in lattice constants, leading to potential microcracks in the crystal. Notably, crystals of Cd$_3$As$_2$ can only be grown at temperatures above 425$^\circ$C, regardless of the growth method.

The complexity of the crystal lattice directly impacts the physical properties of Cd$_3$As$_2$, as well as how we understand them. For instance, the large number of unit cell atoms may
partly complicate {\em ab initio} calculations of the electronic band structure. The complex crystal lattice, with a number of cadmium vacancies, may also be susceptible to small changes in ordering, possibly impacting distinct details in the electronic band structure. Some of the existing controversies about the electronic bands in Cd$_3$As$_2$ may thus stem from differences in the investigated samples' crystal structure, currently prepared using a wide range of crystal growth methods. The crystal lattice complexity is also directly reflected in the optical response of Cd$_3$As$_2$, characterised by a large number of infrared and Raman-active phonon modes~\cite{JandlJRS84,NeubauerPRB16,HoudeSSC86}.

\begin{figure}[b]
\begin{center}
\begin{tabular}{c} 
\includegraphics[width=0.46\textwidth,trim={0cm 0cm 0cm 0cm}]{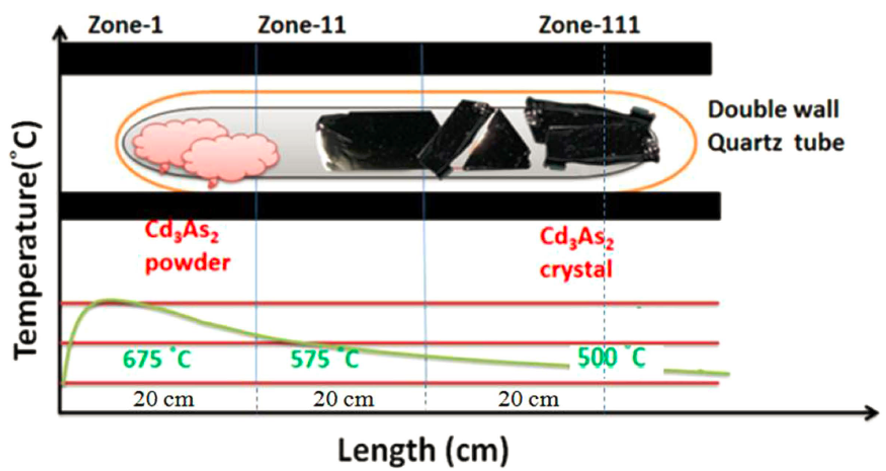}
\end{tabular}
\end{center}
\caption[example]
{\label{SSVG} Schematic cross-sectional side view of an alumina furnace used for self-selecting vapor growth of a Cd$_3$As$_2$ single crystal~\cite{SankarSR15}. The positions of the ampoule in which
transport takes place and the temperature profile of the furnace for Cd$_3$As$_2$ growth are shown. Reprinted from~\cite{SankarSR15}.}
\end{figure}

\section{Technology of sample growth}

The technology of Cd$_3$As$_2$ growth encompasses a broad range of methods dating back over 50 years.
These provide us with a multiplicity of sample forms: bulk or needle-like crystals, thin films, microplatelets, and nanowires. Not only mono- or polycrystalline samples exist, but also amorphous~\cite{ZdanowiczARMS75} samples, all exhibiting different quality and doping levels.
In the past, various techniques have been used, such as the Bridgman~\cite{RogersJPD71} and Czochralski~\cite{SilveyJES61,HiscocksJMS69} methods, sublimation in vacuum or in a specific atmosphere~\cite{Zdanowiczpss64,PawlikowskiTSF75}, pulsed-laser deposition~\cite{DubowskiAPL84}, and directional crystallisation in a thermal gradient~\cite{RosenbergJAP59}.
The results of these techniques were summarised in review articles dedicated to the growth of II$_3$V$_2$ materials~\cite{ArushanovPCGC80,ArushanovPCGC92}.

\begin{figure} [t]
\begin{center}
\begin{tabular}{c} 
\includegraphics[width=0.45\textwidth,trim={0cm 0cm 0cm 0cm}]{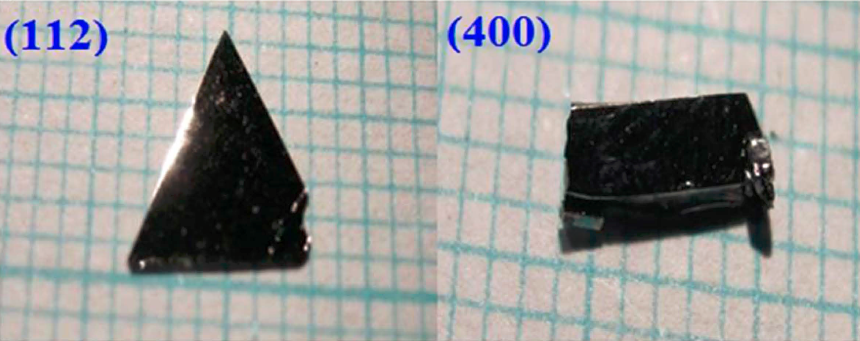}
\end{tabular}
\end{center}
\caption[example]
{\label{OM} Cd$_3$As$_2$ single crystals grown by the SSVG method. The larger facets are (112) and (even 00) planes, reaching an area up to 0.7 cm$^2$. Reprinted from~\cite{SankarSR15}.}
\end{figure}

Most recently, the fast developing field of Cd$_3$As$_2$ has been largely dominated by experiments~\cite{LiangNatureMater14,JeonNatureMater14,BorisenkoPRL14,NeupaneNatureComm14,LiangPRL17,MollNature16,AkrapPRL16} performed on samples prepared by either growth from a Cd-rich melt~\cite{AliIC14}, or self-selecting vapor growth (SSVG)~\cite{SankarSR15}, which are below described in greater detail.
With Cd-rich melt, Cd$_3$As$_2$ is synthesized from a Cd-rich mixture of elements sealed in an evacuated quartz ampoule, heated to 825$^\circ$C, and kept there for 48 hours. After cooling to 425$^\circ$C at a rate of 6 $^\circ$C/h, Cd$_3$As$_2$ single crystals with a characteristic pseudo-hexagonal (112)-oriented facets are centrifuged from the flux.

The SSVG method comprises several steps~\cite{SankarSR15}. In the first, the compound is synthesized from a stoichiometric mixture of elements in an evacuated sealed ampoule, and heated for 4 hours 50 $^\circ$C above the Cd$_3$As$_2$ melting point. The resulting ingot is then purified using a sublimation process in an evacuated closed tube (kept around 800 $^\circ$C), refined with small amounts of excess metal or chalcogen elements (also at 50 $^\circ$C above the melting point), water quenched, annealed (around 700 $^\circ$C) and subsequently air cooled.
Afterwards, the ingot is crushed and then sieved (targeting a particle size of 0.1-0.3~mm), with the obtained precursors then sealed in an evacuated ampoule. This is then inserted for about 10 days into a horizontal alumina furnace, whose temperature profile is shown in Fig.~\ref{SSVG}.
The resulting crystals are plate-like or needle-like monocrystals, or polycrystals with large grains. The crystals have primarily (112)-oriented, but occasionally also ($n$00)-oriented facets, with $n$ an even number, see Fig.~\ref{OM}.

\begin{figure}
	\includegraphics[width=0.4\textwidth,trim={1cm 1cm 1cm 1cm}]{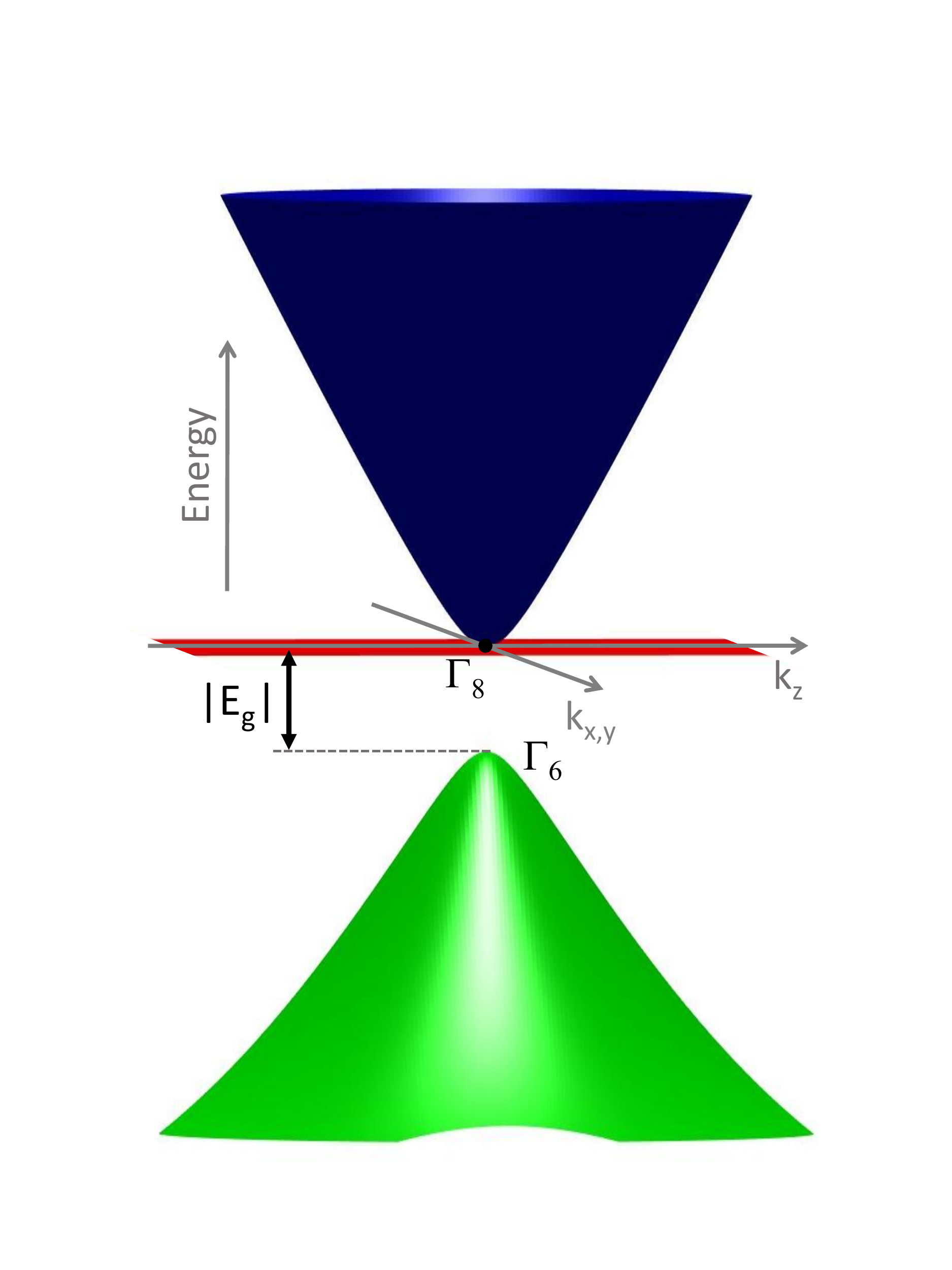}
	\caption{The electronic band structure of a cubic semiconductor/semimetal implied by the Kane mode in the vicinity of the
Brillouin zone center. At energies significantly larger than the band gap, both conduction and valence bands exhibit nearly
conical shapes. For a negative band gap $E_g$ (chosen in the plotted case), the system is characterized by
an inverted ordering of bands, and closely resembles the well-known semimetal HgTe~\cite{WeilerSaS81}. A positive band gap would imply a band structure typical of narrow-gap semiconductors such as InSb~\cite{CardonaYu}. \label{Kane}}
\end{figure}

Crystals prepared using the two methods discussed above, as well as past methods, display $n$-type conductivity with a relatively high density of electrons.
As-grown and without specific doping, the crystals rarely show an electron density below $10^{18}$~cm$^{-3}$. This  translates into typical Fermi levels in the range of $E_F=100-200$~meV measured form the charge-neutrality point.
In literature this rather high doping is usually associated with the presence of arsenic vacancies~\cite{SpitzerJAP66}. The doping of Cd$_3$As$_2$ has been reported to vary with conventional annealing~\cite{RamboCJP79}. It seems to decrease after thermal cycling between
room and helium temperatures~\cite{CrasseePRB18}. A combined optical and transport study also revealed a relatively thick (10-20~$\mu$m) depleted layer
on the surface of Cd$_3$As$_2$ crystals~\cite{SchleijpeIJIMW84}. Optical studies also show relatively large inhomogeneities of up to 30\% in the electron density, at the scale of hundred microns, in crystals prepared using different growth methods, with $x$-ray studies indicating the presence of systematic twinning~\cite{CrasseePRB18}.

Recently, there has been progress in other growth methods. These include the CVD technique, employed to fabricate Cd$_3$As$_2$ nanowires~\cite{WangNC16}.
Other methods, such as pulsed laser deposition in combination with solid phase epitaxy~\cite{UchidaNC17} and molecular beam epitaxy~\cite{YuanNL17,SchumannPRL18},
have been successfully employed to prepare Cd$_3$As$_2$ layers with a thickness down to the nanometer scale. Most importantly, when the thickness of Cd$_3$As$_2$ films is reduced down to the tens of nanometers, ambipolar gating of Cd$_3$As$_2$  becomes possible~\cite{GallettiPRB18}.

\section{Electronic band structure: theoretical views}

Soon after the first experimental studies appeared~\cite{RosenbergJAP59,Zdanowiczpss64,Haidemenakis66,Sexerpss67,RosenmanJPCS69}, the band structure of Cd$_3$As$_2$ was approached theoretically. In this early phase, the electronic band structure
was described using simple effective models, developed in the framework of the standard $k.p$ theory. Such models were driven by
the similarity between the crystal lattice -- and presumably also electronic states -- in Cd$_3$As$_2$, and conventional
binary semiconductors/semimetals such as GaAs, HgTe, CdTe or InAs~\cite{CardonaYu}.
In all of these materials, the electronic bands in the vicinity to the Fermi energy are overwhelmingly composed of cation $s$-states and anion $p$-states.
In the case of Cd$_3$As$_2$, those are cadmium-like cations and arsenic-like anions.

In the first approach, the electronic band structure of Cd$_3$As$_2$ has been described using the conventional
Kane model~\cite{KaneJPCS57}, which is widely applied in the field of zinc-blende semiconductors and which has also been successfully expanded
to account for effects due to quantizing magnetic fields~\cite{BowersPRB59,PidgeonPR66}.
To the best of our knowledge, the very first attempt to interpret the experimental data collected on Cd$_3$As$_2$ using the Kane model was presented by Armitage and Goldsmid~\cite{ArmitagePLA68} in 1968. Later on, similar studies appeared~\cite{RogersJPD71,CaronPRB77,AubinPRB77,Jay-GerinJLTP77} but with no clear consensus on the particular band structure parameters.

The size and nature of the band gap, describing the separation between the $p$-like and $s$-like states at the $\Gamma$ point, remained a main source of controversy~\cite{ZdanowiczARMS75}. Most often, relatively small values of either inverted or non-inverted
gaps have been reported, based on the analysis of different sets of experimental data~\cite{ArmitagePLA68,RosenmanJPCS69,CaronPRB77}.
Notably, when the band gap is significantly smaller than the overall energy scale of the considered band structure, the Kane model implies approximately
conical conduction and valence bands, additionally crossed at the apex by a relatively flat band (Fig.~\ref{Kane})~\cite{Kacmanppsb71,OrlitaNaturePhys14}.
This band is usually referred to as being heavy-hole-like in semiconductor physics, and may be considered as flat only in the vicinity of the $\Gamma$ point.
At larger momenta, it disperses with a characteristic effective mass close to unity.

\begin{figure}[t]
	\includegraphics[width=0.4\textwidth,trim={1cm 1cm 1cm 1cm}]{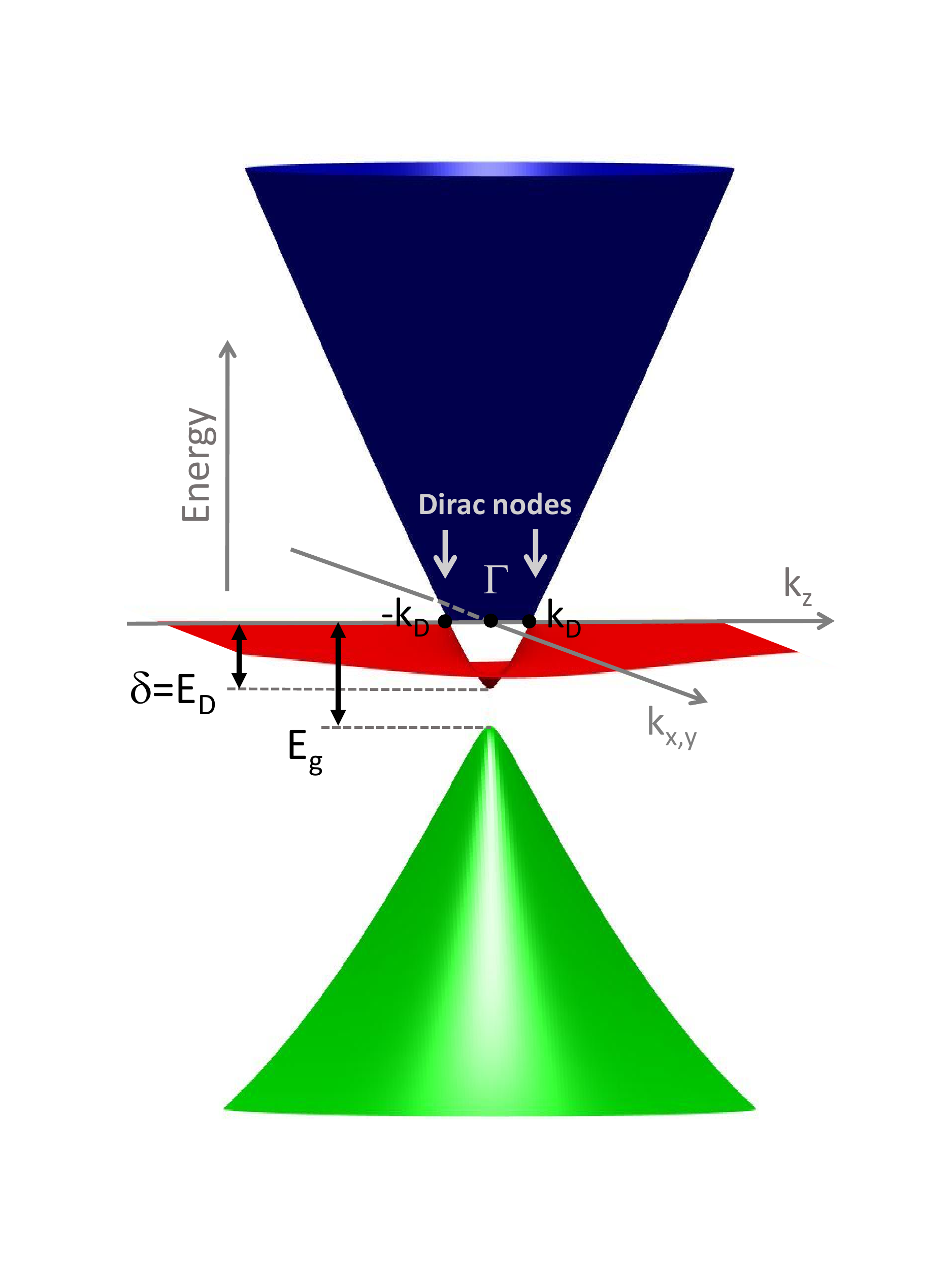}
	\caption{The schematic view of the electronic band structure of Cd$_3$As$_2$ proposed by Bodnar~in Ref.~\cite{Bodnar77}. Three electronic bands form two types of 3D
conical structures: a single cone hosting Kane electrons at the large energy scale, appearing due to the vanishing band gap, and two
highly tilted and anisotropic 3D Dirac cones at low energies. These are marked by vertical gray arrows and emerge due to the partly avoided crossing of the flat band with the conduction band. The energy scale of massless Dirac electrons $E_D$ exactly matches the size of crystal field splitting parameter $\delta$.
\label{Bodnar}}
\end{figure}

Importantly, the presence of a 3D conical band in the Kane model is a result of an approximate accidental degeneracy of $p$-like and $s$-like states at the $\Gamma$ point, therefore making this cone not protected by any symmetry. Additionally, this cone is described
by the Kane Hamiltonian~\cite{Kacmanppsb71}, which is clearly different from the Dirac Hamiltonian. It should therefore not be confused with
3D Dirac cones subsequently predicted for Cd$_3$As$_2$~\cite{WangPRB13}. The term \emph{massless Kane electron} was recently introduced~\cite{OrlitaNaturePhys14,MalcolmPRB15,TeppeNatureComm16,AkrapPRL16} to distinguish those two types of 3D massless charge carriers. Therefore, at the level of the strongly
simplifying Kane model, Cd$_3$As$_2$ does not host any massless 3D Dirac electrons. A nearly conical band, the presence of which was deduced from early transport
experiments indicating energy-dependent effective mass~\cite{ArmitagePLA68,RosenmanJPCS69} (Fig.~\ref{Rosenman}), was then interpreted in terms of the Kane model assuming a nearly vanishing band gap.

\begin{figure*}[t]
	\includegraphics[width=1\textwidth,trim={0cm 0cm 0cm 0cm}]{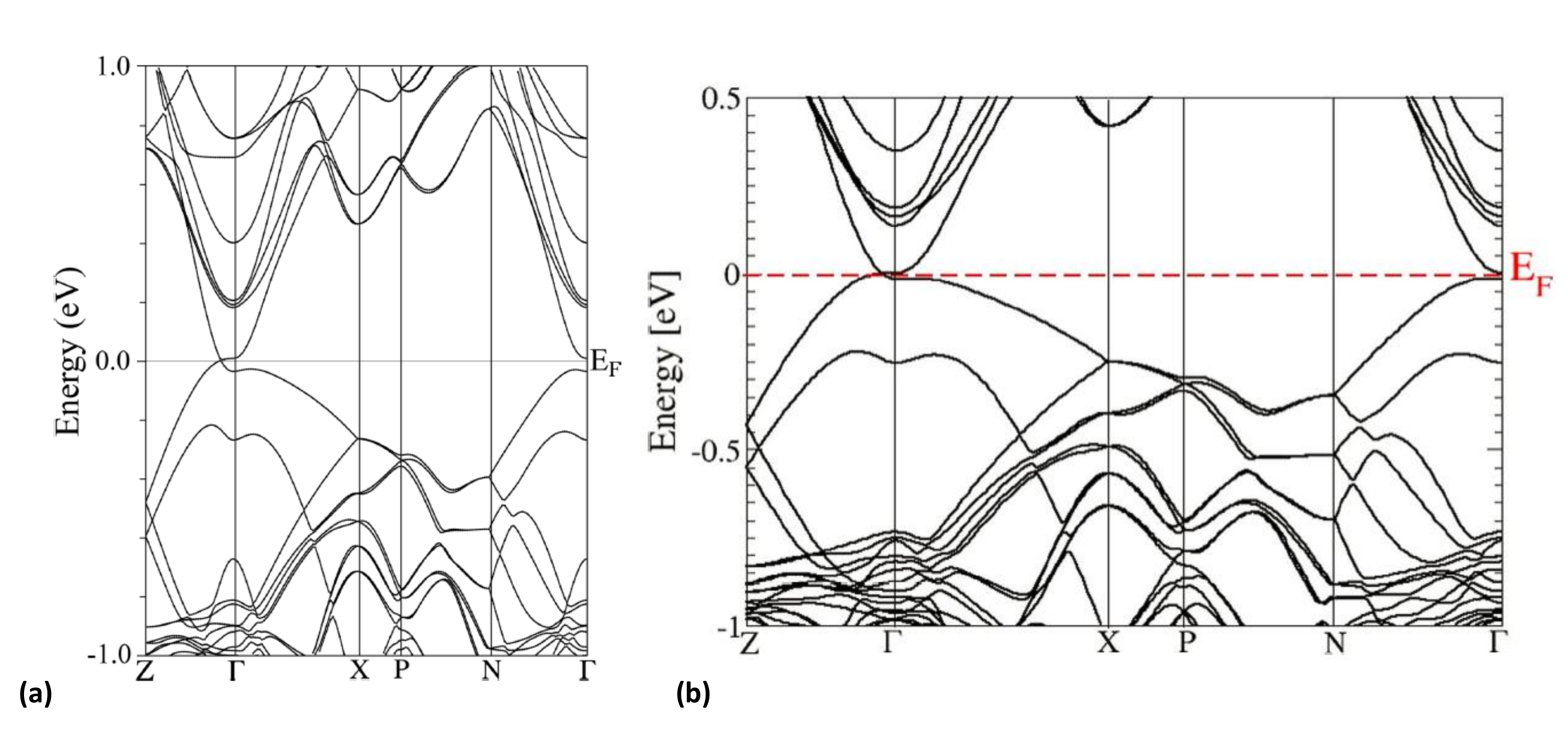}
\caption{The band structure of Cd$_3$As$_2$ deduced theoretically using \emph{ab initio} approach by Ali~\emph{et al.}~\cite{AliIC14} and Conte~\emph{et al.}~\cite{ConteSR17},
using GGA approximations, in parts (a) and (b), respectively. In both cases, the 3D Dirac cones appear at the Fermi energy in close vicinity to the $\Gamma$ point,
due the symmetry-allowed crossing of bands along the $\Gamma$-Z line. The energy scale of the massless Dirac electrons $E_D$ in these two cases may be estimated to be 45 and 20~meV, respectively.
Part (a) reprinted with permission from~\cite{AliIC14}, copyright (2014) American Chemical Society. Part (b) reprinted from~\cite{ConteSR17}.
\label{Ab initio}}
\end{figure*}

An improved effective model, which takes into account the tetragonal nature of Cd$_3$As$_2$, not included in the conventional Kane model
for zinc-blende semiconductors, has been proposed by Bodnar~\cite{Bodnar77}. The tetragonal distortion of a nearly cubic lattice lifts the degeneracy of $p$-type states (light and heavy hole bands) at the $\Gamma$ point; this degeneracy is typical of all zinc-blende semiconductors~\cite{CardonaYu}.
A closer inspection of the theoretical band structure reveals two specific points at the tetragonal axis. At these points the conduction and flat heavy-hole valence bands
meet and form two highly anisotropic and tilted cones (Fig.~\ref{Bodnar}). These may be associated with symmetry-protected 3D Dirac cones. Nevertheless, it was not until the work by Wang~\emph{et al.}~\cite{WangPRB13} that such a prediction appeared explicitly in the literature. The crystal-field splitting parameter $\delta$~\cite{KildalPRB74} is employed by Bodnar~\cite{Bodnar77} to quantify the impact of the tetragonal
distortion of the cubic lattice. This parameter directly corresponds to the energy scale $E_D$ of the symmetry-protected massless Dirac electrons in Cd$_3$As$_2$ (cf. Figs.~\ref{Scheme} and~\ref{Bodnar}).
The work of Bodnar also inspired other theorists, who used the proposed model for electronic band structure calculations in quantizing magnetic fields~\cite{Wallacepss79,SinghJPC83,SinghJPC84}.

The above described effective models have been applied with some success to explain experimental data available those times. However, these models should be confronted with the much broader experimental knowledge acquired recently. Similar to all other models based on the $k.p$ expansion, their validity is limited to the near vicinity of the Brillouin zone center, and to the number of spin-degenerate bands taken into account
(restricted to 4 in the Kane/Bodnar model). Moreover, the number of band structure parameters is strongly restricted in these models. This ensures the models' relative
simplicity, but at the same time, limits the potential to describe the band structure in greater detail, even in the immediate vicinity of the $\Gamma$ point.

It is therefore important to reconcile the implications of such effective models with other theoretical approaches. Such band structure calculations already appeared in the early stages of research on Cd$_3$As$_2$. These were based on either pseudopotential calculations~\cite{Lin-ChungPRB69,Lin-Chungpss71,Dowgiallo-Plenkiewiczpps79} or the
semiempirical tight-binding method~\cite{SieranskiPRB94}. More recently, {\em ab initio} calculations have been performed, with their main focus on the Dirac-like states. To the best of our knowledge, the first {\em ab initio} study of Cd$_3$As$_2$ predicting the presence of 3D massless Dirac electrons was presented by Wang~\emph{et al.}~\cite{WangPRB13}.
Nevertheless, the considered space groups did not comprise the most probable one, I4$_1$/acd (No.~110)~\cite{PietraszkoPSS73,AliIC14}. Other {\em ab initio} calculations may be found in Refs.~\cite{LiuNatureMater14,NeupaneNatureComm14,BorisenkoPRL14,ConteSR17,ShchelkachevIM18}, often carried out in support of experimental findings.

Even though the results of {\em ab initio} calculations may differ in some details -- most likely related to the particular approximation of exchange and correlation functionals and the size of the unit cell considered -- they provide us with a rather consistent theoretical picture of the electronic bands in Cd$_3$As$_2$. They predict that Cd$_3$As$_2$ is a semimetal, with a pair of well-defined 3D Dirac
cones, emerging at relatively low energies. In line with symmetry arguments, the Dirac nodes are found to be located at the tetragonal axis, with an exception of Ref.~\cite{LiuNatureMater14}.

\begin{table*}[t]
\caption{\label{Table}
The band structure parameters of Cd$_3$As$_2$: the energy scale $E_D$ (defined in Fig.~\ref{Scheme}), the Dirac point position
$k_D$ and the crystallographic axis at which the Dirac nodes are located, as deduced using different experimental techniques
or theoretical calculations/analysis. The velocity parameter is usually found to be around, or slightly below, $10^6$~m/s in the direction perpendicular
to the line connecting Dirac nodes, and reduced down to the $10^5$~m/s range along this direction.}
\begin{ruledtabular}
\begin{tabular}{ l c c r}
\textrm{Technique} & \textrm{Scale} $E_D$ (meV) & \textrm{Location} $k_D$ (nm$^{-1}$) & \textrm{Orientation}  \\
\colrule
\emph{ab initio}, GGA, I4$_1$cd~\cite{WangPRB13} & 40 & 0.32 & [001]\\
\emph{ab initio}, GGA, I4$_1$/acd~\cite{AliIC14} & 45 & 0.4 & [001]\\
\emph{ab initio}, GGA, I4$_1$/acd~\cite{ConteSR17} & 20& 0.23 & [001]\\
\emph{ab initio}, GGA, I4$_1$cd~\cite{LiuNatureMater14} & 200 & 1.2 & [112]\footnote{equivalent to the [111] direction when the cubic cell approximation is considered as in Ref.~\cite{LiuNatureMater14}}\\
ARPES~\cite{LiuNatureMater14} & several hundred & 1.6 & [112]\footnotemark[1]\\
ARPES~\cite{NeupaneNatureComm14} & several hundred &  & [001]\\
ARPES~\cite{BorisenkoPRL14} & several hundred & & [001] \\
STM/STS~\cite{JeonNatureMater14} & 20 & 0.04 & [001]\\
Magneto-optics \cite{HaklPRB18,AkrapPRL16} & $<$40 & $<$0.05 & [001]\\
Magneto-transport \& Bodnar model~\cite{RosenmanJPCS69,Bodnar77} & 85 & 0.15 & [001]\\
Magneto-transport \cite{ZhaoPRX15} & $<$200 &  & [001]\\
\end{tabular}
\end{ruledtabular}
\end{table*}

When the non-centrosymmetric I4$_1$cd space group is considered, the double degeneracy of the Dirac cones due to spin may be lifted~\cite{WangPRB13}. The loss of inversion symmetry then transforms the 3D Dirac semimetal into a 3D Weyl semimetal with two pairs of Weyl nodes. The Dirac cones are dominantly formed from $p$-type arsenic states, and are well separated from the bands lying at higher or lower energies.
Notably, the $s$-like cadmium states are found well below the Fermi energy. The band structure is therefore inverted and may be formally described by a negative band gap ($E_g\approx-0.5$~eV). To certain extent, the electronic bands in Cd$_3$As$_2$ resemble those in HgTe, which is another semimetal with an inverted band structure~\cite{WeilerSaS81}.

To illustrate the typical results of {\em ab initio} calculations, two examples have been selected from Refs.~\onlinecite{AliIC14} and \onlinecite{ConteSR17} and
plotted in Figs.~\ref{Ab initio}a and b, respectively. The 3D Dirac cones are located in close vicinity to the $\Gamma$ point, with the charge neutrality (Dirac) points
at the $\Gamma$-Z line and the Fermi level. The parameters of these cones derived in selected {\em ab initio} studies are presented
in Tab.~\ref{Table}. For instance, Conte~\emph{et al.}~\cite{ConteSR17} deduced, using the GGA approximation, that the energy scale of massless Dirac electrons reaches
$E_D\approx 20$~meV, the strongly anisotropic velocity parameter is of the order of $10^5$~m/s, and two Dirac nodes
are located at $k_D=\pm 0.23$~nm$^{-1}$.

Let us now explore the experiments carried out on Cd$_3$As$_2$ using various techniques. These provided us with unambiguous
evidence for the conical features in this material. However, let us clearly note from the beginning that the scale of the conical bands deduced -- using ARPES~\cite{LiuNatureMater14,NeupaneNatureComm14,BorisenkoPRL14} and STM/STS~\cite{JeonNatureMater14} techniques, or from the optical response~\cite{AliIC14,NeubauerPRB16,AkrapPRL16} -- is not consistent with the energy scale predictions for massless Dirac electrons based on {\em ab initio} calculations. While the theoretically expected scale of
massless Dirac electrons rarely exceeds $E_D\sim 100$~meV (Tab.~\ref{Table}), the experimentally observed cones extend over a significantly broader interval of energies.

In the following sections, we employ, for the sake of brevity, a simplified notation and refer to the Kane and Dirac models, which both may -- at least from the theoretical viewpoint -- explain the presence of massless electrons in Cd$_3$As$_2$. In the case of the Dirac model, we always consider 3D massless electrons, which are described by the Dirac equation with the zero rest mass and the presence of which is protected by the discrete ($C_4$) rotational symmetry~\cite{WangPRB13}. In the case of Kane model, zero or vanishing band gap is implicitly assumed.

\section{Angular-resolved photoemission spectroscopy}

The ARPES technique provided us with a solid piece of evidence for conical features in the electronic band structure
of Cd$_3$As$_2$, soon after predictions of Dirac-like states by Wang~\emph{et al.}~\cite{WangPRB13}. This largely contributed to the renewed interest in the electronic properties of this material. Such initial observations were made by several groups~\cite{LiuNatureMater14,BorisenkoPRL14,NeupaneNatureComm14,YiSR14}, and elaborated further later on~\cite{RothPRB18}.

\begin{figure}
	\includegraphics[width=0.35\textwidth]{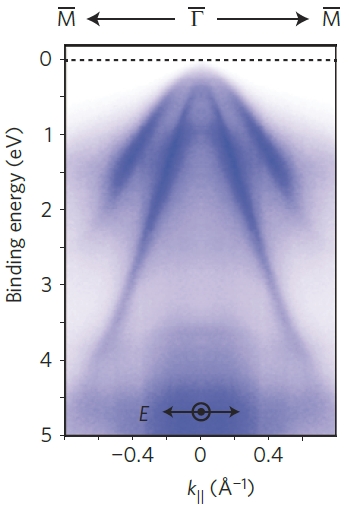}[b]
	\caption{The valence bands in Cd$_3$As$_2$ visualized by low-temperature ARPES technique by Liu~\emph{et al.}~\cite{LiuNatureMater14}. The data was collected on the (112)-terminated
               surface. The widely extending 3D conical band -- characterized by a velocity parameter of $1.3\times10^6$~m/s and interpreted in terms of 3D massless Dirac electrons -- coexists with another hole-like weakly dispersing parabolic band, also observed in Ref.~\cite{NeupaneNatureComm14}.
               Reprinted by permission from Springer Nature: Nature Materials~\cite{LiuNatureMater14}, copyright (2014). \label{ARPES-Liu}}
\end{figure}

\begin{figure}[b]
	\includegraphics[width=0.35\textwidth]{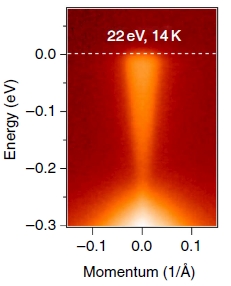}
	\caption{The conduction band of Cd$_3$As$_2$ visualized by ARPES by Neupane~\emph{et al.}~\cite{NeupaneNatureComm14}. The data were collected on the (001)-terminated surface of  Cd$_3$As$_2$ at 14~K and show the dispersion in the direction perpendicular to the $\Gamma$-Z line. The observed 3D conical band was interpreted in terms of 3D massless Dirac electrons, implying a velocity parameter of $1.5\times10^6$~m/s. Reprinted by permission from Springer Nature: Nature Communications \cite{NeupaneNatureComm14}, copyright (2014). \label{ARPES-Neupane}}
\end{figure}

Characteristic data collected in ARPES experiments on Cd$_3$As$_2$~\cite{LiuNatureMater14,NeupaneNatureComm14} are plotted in Figs.~\ref{ARPES-Liu} and \ref{ARPES-Neupane},
and respectively show well-defined conical features for both valence and conduction bands.
The conical features in Fig.~\ref{ARPES-Liu} and \ref{ARPES-Neupane} were interpreted in terms of bulk states.  Liu~\emph{et al.}~\cite{LiuNatureMater14} and Neupane~\emph{et al.}~\cite{NeupaneNatureComm14} concluded the presence of a pair of 3D Dirac nodes at the [112] and [001] axes, respectively. However, it is worth noting that the orientation along the [112] axis -- or alternatively, along the [111] axis when an approximately cubic unit cell is considered like it is the case in Ref.~\cite{LiuNatureMater14} -- is not consistent with expectations based on symmetry arguments~\cite{YangNatureComm14}, which only allow the Dirac nodes to be present at the tetragonal axis (the [001] direction).
The velocity parameter was found to be close to $10^6$~m/s in the plane perpendicular to the axis connecting the Dirac nodes, and reduced down to $3\times 10^5$~m/s~\cite{LiuNatureMater14} along this axis.
The ARPES data in Refs.~\cite{LiuNatureMater14,NeupaneNatureComm14} do not directly show any signatures of Dirac cones merging via the corresponding Lifshitz points. Nevertheless,
the indicated velocity parameters, and the position of the cones ($k_D$), allow us to estimate the scale of massless Dirac electrons $E_D$ to be several hundred meV or more.

ARPES data similar to Refs.~\cite{LiuNatureMater14,NeupaneNatureComm14}, obtained on the [112]-terminated surface of Cd$_3$As$_2$, were also presented by Borisenko~\emph{et al.}~\cite{BorisenkoPRL14}, who primarily focused on the conical feature in the conduction band. The presence of a pair of symmetry-protected 3D Dirac cones, with the corresponding nodes
at the [001] axis, has been concluded, and the electron velocity parameter $v\approx0.8\times10^6$~m/s deduced.
The Fermi energy in the studied $n$-doped sample exceeded $E_F\approx 200$~meV, and may serve as a lower bound for the $E_D$ parameter.
Since the shape of the observed conical band does not provide any signature of the approaching upper Lifshitz point, one may conclude that $E_D\gg E_F$.

The basic parameters of Dirac-like conical bands deduced from the above cited ARPES experiments have been compared in Tab.~\ref{Table} with results of other experimental techniques, which are discussed in detail later on. One may see that there is a considerable spread of the reported values for both the energy scale, and the position of the Dirac cones in Cd$_3$As$_2$.
For instance, the energy scale consistent with the ARPES data exceeds by almost two orders of magnitude the estimate based on STM/STS measurements~\cite{JeonNatureMater14}.
At present, the reason for this disagreement remains unclear.

More recently, other ARPES experiments on Cd$_3$As$_2$ have been performed by Roth~\emph{et al.}~\cite{RothPRB18}. They provided experimental data similar to previous studies, but differed in their interpretation. From their experiment realized on a sample with a (112)-terminated surface, they concluded that
a part of observed conical features does not come from bulk, but instead originate from the surface states (cf. Ref.~\cite{YiSR14}).
Let us note that, since each Dirac node is composed of two Weyl nodes with opposite chiralities~\cite{PotterNatureComm14}, such surface states may have the form of the Fermi arcs in a 3D symmetry-protected Dirac semimetal.

To the best of our knowledge, the electronic band structure observed in ARPES experiments has not yet been in detail compared with expectations of the Kane/Bodnar model.
Nevertheless, it is interesting to note that the conical valence band observed in the ARPES data (Fig.~\ref{ARPES-Liu}) coexists with another weakly dispersing hole-like parabolic band~\cite{NeupaneNatureComm14}, which has a characteristic effective mass close to unity. Such a band is not expected in the model of 3D massless Dirac electrons, nevertheless,
it could be straightforwardly explained within the Kane/Bodnar picture with a small or vanishing band gap.

\section{Scanning tunneling spectroscopy and microscopy}

Similar to ARPES, the STS/STM technique also played an important role in the recent revival of Cd$_3$As$_2$.
The data collected in STS experiments performed in magnetic fields~\cite{JeonNatureMater14} revealed, via the characteristic $\sqrt{B}$-dependence of Landau levels, the presence of a single widely extending conical band located at the center of the Brillouin zone. In the literature, this observation is often taken as experimental evidence
for the symmetry-protected 3D massless Dirac electrons in Cd$_3$As$_2$. However, according to Jeon~\emph{et al.}~\cite{JeonNatureMater14}, the existence of ``this extended linearity is not guaranteed by the Dirac physics around the band inversion''.

Such a conclusion agrees with theoretical expectations based on symmetry arguments.
These exclude the existence of a symmetry-protected Dirac cone located at the center of the Brillouin zone~\cite{YangNatureComm14}. In addition, this observation is in line with the Kane/Bodnar models used to explain the band structure of Cd$_3$As$_2$ in the past. Nevertheless, Jeon~\emph{et al.}~\cite{JeonNatureMater14} also conclude -- by extrapolating their Landau level spectroscopy data to vanishing magnetic fields -- that a pair of
symmetry-protected Dirac cones emerges at low energies. They give a rough estimate of $E_D\approx20$~meV for the characteristic Dirac energy scale.

Beyond insights into the bulk electronic states of Cd$_3$As$_2$, the natural sensitivity of the STM/STS technique to the surface of explored systems
may provide us with deeper knowledge about their surface states. A 3D symmetry-protected Dirac semimetal like Cd$_3$As$_2$ can be viewed as two copies of a 3D Weyl semimetal with nodes having opposite chiralities; two sets of Fermi arcs are expected on the surface, see Ref.~\cite{KargarianPNAS16} for details. The recent STS/STM study~\cite{ButlerPRB17} dedicated to the surface reconstruction of cadmium vacancies in Cd$_3$As$_2$  may be considered as the initial step in such investigations.

\begin{figure}[t]
	\includegraphics[width=0.37\textwidth,trim={1cm 1cm 1cm 1cm}]{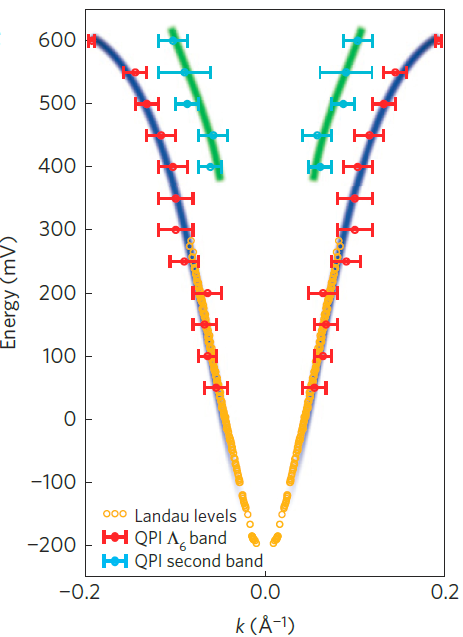}
	\caption{The approximately conical band centered at the $\Gamma$ point deduced from Landau-level spectroscopy (open circles) and quasi-particle interference (QPI) pattern (closed circles with error bars) in the STS/STM experiments performed by~Jeon~\emph{et al.}~\cite{JeonNatureMater14}. The slope of the conical band corresponds to a velocity parameter of $0.94\times10^6$~m/s. Reprinted by permission from Springer Nature: Nature Materials~\cite{JeonNatureMater14}, copyright (2014). \label{Jeon-STS}}
\end{figure}

\section{Optical properties}
The optical and magneto-optical properties of Cd$_3$As$_2$ have been a topic of study for more than 50 years, and resulted in a series of works~\cite{TurnerPR61,Haidemenakis66,ZdanowiczPSS67}. These were often interpreted in terms of Kane/Bodnar models~\cite{WagnerJPCSS71,RogersJPD71,RadoffPRB72,AubinPRB77,
GeltenSSC80,AubinPRB81,Jay-GerinSSC83,SinghPB83,SinghJPC83,HoudeSSC86,LamraniCJP87,AkrapPRL16,HaklPRB18,CrasseePRB18}, and more recently using the picture of 3D massless Dirac electrons~\cite{NeubauerPRB16,JenkinsPRB16,YuanNL17,UykurPRB18}.

Early optical studies were focused on the basic character of the electronic band structure in Cd$_3$As$_2$.
They aimed at clarifying the existence of a band gap and at determining its size. Using infrared reflectivity and transmission techniques, Turner~\emph{et al.}~\cite{TurnerPR61} concluded that Cd$_3$As$_2$ should be classified as a narrow-gap semiconductor with a direct band gap of 0.16~eV.
Similarly, an indirect band gap around 0.2~eV, was concluded by~Zdanowicz~\emph{et al.}~\cite{ZdanowiczPSS67} based on transmission experiments.
Much lower values were found in magneto-optical studies. Haidemenakis~\emph{et al.}~\cite{Haidemenakis66} concluded $E_g<30$~meV, suggesting a semimetallic nature of Cd$_3$As$_2$. The difference in conclusions between optical and magneto-optical studies may be related to the large doping
of the explored samples. In that case, the onset of interband absorption, often referred to as the optical band gap,
appears due to the Moss-Burstein shift (Pauli blocking)~\cite{BursteinPR54}, at photon energies significantly exceeding the size of the energy band gap.

Further series of optical and magneto-optical Cd$_3$As$_2$ studies were performed on various mono- or polycrystalline samples during the seventies and eighties. The collected data was primarily analyzed within the framework of the Kane model~\cite{WagnerJPCSS71,RogersJPD71,RadoffPRB72,AubinPRB77,GeltenSSC80}, and later on the Bodnar model~\cite{AubinPRB81,Jay-GerinSSC83,LamraniCJP87}. The authors of these works concluded that the electronic band structure at the $\Gamma$ point is fairly well described using these models: it is inverted, with the $p$-type arsenic states above the $s$-like cadmium states, and characterized by a relatively small negative band gap. A schematic view of such a band structure is plotted in Figs.~\ref{Kane} and \ref{Bodnar}. The deduced value of the band gap reached $E_g \approx -0.2$~eV~\cite{WagnerJPCSS71,RadoffPRB72,AubinPRB77} at room or low temperatures, but also values as low as $E_g \approx -0.1$~eV have been reported~\cite{AubinPRB81,LamraniCJP87,Jay-GerinSSC83}. Particular attention has been paid to the profile and position of the weakly dispersing band, which is nearly flat in the vicinity of the $\Gamma$ point. Several times, the possibility of the band maxima located at non-zero momenta has been discussed~\cite{AubinPRB77, GeltenSSC80}. At the same time, Lamrani and Aubin concluded a surprisingly flat heavy-hole band, when the theoretically expected Landau-quantized Bodnar band structure~\cite{SinghPB83,SinghJPC83} was confronted
with the experimentally determined energies of interband inter-Landau level excitations~\cite{LamraniCJP87}.

\begin{figure}
	\includegraphics[trim = 0mm 0mm 0mm 0mm, clip=true, width=1\columnwidth]{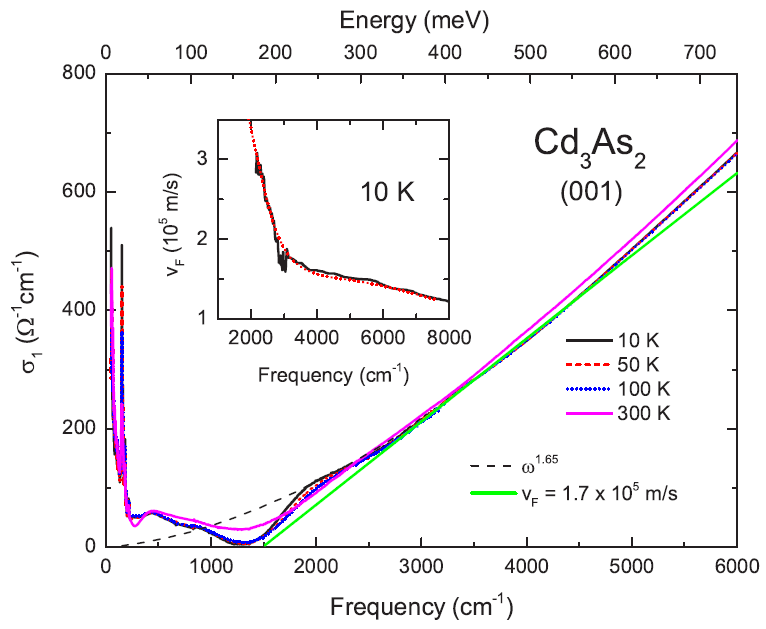}
	\caption{The real part of the optical conductivity of Cd$_3$As$_2$ obtained by Neubauer~et al.~\cite{NeubauerPRB16}. At low energies, the conductivity is dominated by Drude (free-carrier) response accompanied by a series of infrared active phonon modes. Above the onset of the interband absorption around $1500-2000$~cm$^{-1}$, the optical conductivity increases roughly linearly with the photon energy, which is behavior expected for massless charge carries in 3D.
Reprinted with permission from~\cite{NeubauerPRB16}, copyright (2016) by the American Physical Society.\label{Neubauer}}
\end{figure}

The renewed interest in Cd$_3$As$_2$ motivated several groups to take a fresh look at the optical, magneto-optical, and ultra fast optical properties of this material~\cite{WeberAPL15,NeubauerPRB16,JenkinsPRB16,AkrapPRL16,YuanNL17,SharafeevPRB17,HaklPRB18,CrasseePRB18,WeberAPL15,ZhuAPL15,LuPRB17,LuPRB18}. Thanks to this, to the best of our knowledge, the optical conductivity of Cd$_3$As$_2$ (Fig.~\ref{Neubauer}) has been extracted from the experimental data for the very first time~\cite{NeubauerPRB16}. At low energies,
the optical conductivity is characterized by a pronounced Drude peak due to the presence of free charge carriers, and a rich set of phonon excitations. Similar to the Raman response~\cite{JandlJRS84,SharafeevPRB17}, the complexity of the phonon-related response directly reflects the relatively high number of atoms in the Cd$_3$As$_2$ unit cell.
Above the onset of interband absorption, optical conductivity increases with a slightly superlinear dependence on photon energy. Such behavior is not far from the expectation for
3D massless charge carriers, $\sigma(\omega)\propto \omega$~\cite{GoswamiPRL11,TimuskPRB13}. The observed optical response was thus interpreted in terms of 3D massless Dirac electrons,
with a low anisotropy and a velocity parameter lying in the range of $1.2-3.0\times10^{5}$~m/s. No clear indications of Dirac cones merging at the Lifshitz points were found. Reflectivity data similar to Ref.~\cite{NeubauerPRB16} were also presented by Jenkins~\emph{et al.}~\cite{JenkinsPRB16}, with basically the same conclusions.

The Dirac model was similarly used to interpret the classical-to-quantum crossover of cyclotron resonance observed in
magneto-transmission data collected on thin MBE-grown Cd$_3$As$_2$ layers~\cite{YuanNL17} as well as pump-probe experiments in the visible spectral range, which
revealed the transient reflection~\cite{WeberAPL15} and transmission~\cite{ZhuAPL15}. These latter experiments show that a hot carrier distribution is obtained after 400--500~fs, after which the charge carriers relax by two processes. Subsequent pump-probe experiments using mid-infrared~\cite{LuPRB17} and THz probe~\cite{LuPRB18} confirm this two-process relaxation, which can be qualitatively reproduced using a two temperature model.

\begin{figure}[t]
\includegraphics[trim = 0mm 0mm 0mm 0mm, clip=true, width=0.9\columnwidth]{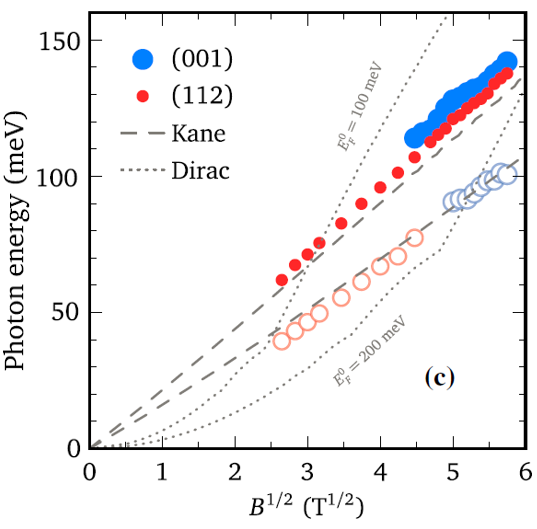}
\caption{\label{CR} The energies of cyclotron resonance excitations observed in Ref.~\cite{AkrapPRL16} as a function of applied magnetic field,
obtained on (112)- and (001)-oriented Cd$_3$As$_2$ samples. The dashed and dotted lines show theoretically expected positions within the Kane
and Dirac models, see Ref.~\cite{AkrapPRL16} for details. Adapted with permission from~\cite{AkrapPRL16}, copyright (2016) by the American Physical Society.}
\end{figure}

A different view was proposed in a recent magneto-reflectivity study of Cd$_3$As$_2$ by Akrap~et~al.~\cite{AkrapPRL16} (see Fig.~\ref{CR}). In high magnetic fields,
when the samples were pushed into their corresponding quantum limits with all electrons in the lowest Landau level,
the $\sqrt{B}$ dependence of the observed cyclotron mode was found to be inconsistent with 3D massless Dirac electrons.
The data was interpreted in terms of the Kane/Bodnar model, with a vanishing band gap.
This model also leads to the appearance of 3D massless electrons, and consequently, to a magneto-optical response linear in $\sqrt{B}$. In this model,
the $\sqrt{B}$ dependence of the cyclotron mode is also expected in the quantum limit. This is in contrast to 3D massless Dirac electrons, which host characteristic so-called zero-mode Landau levels. These are independent of magnetic field, disperse linearly with momentum along the direction of the applied field,
and imply a more complex cyclotron resonance dependence in the quantum limit.  An approximately isotropic velocity parameter was found for
the nearly conical conduction band: $v \approx 0.9\times 10^6$~m/s~\cite{AkrapPRL16}. Estimates of the band gap and crystal field splitting parameter at low temperatures
were obtained in a subsequent magneto-transmission study performed on thin Cd$_3$As$_2$ slabs~\cite{HaklPRB18}: $E_g=-(70\pm20)$ meV and $|\delta|=E_D<40$~meV.

Lately, spatially resolved infrared reflectivity has been used to characterize the homogeneity of Cd$_3$As$_2$ crystals~\cite{CrasseePRB18}.
In all the studied samples, independently of how they were prepared and how they were treated before the optical experiments, conspicuous fluctuations
in the carrier density up to 30\% have been found. These charge puddles have a characteristic scale of 100~$\mu$m. They become more pronounced at low temperatures,
and possibly, they become enhanced by the presence of crystal twinning. Such an inhomogeneous distribution of electrons may be a generic property of
all Cd$_3$As$_2$ crystals, and should be considered when interpreting experimental data collected using other techniques.

\section{Magneto-transport properties and quantum oscillations}

Renewed interest in Cd$_3$As$_2$ brought upon an explosion of interest in magneto-transport studies. Many of these studies focused on the previously reported very high carrier mobility~\cite{RosenbergJAP59}, and its possibilities for device application.
Often, newer transport results are interpreted within the scenario of two 3D Dirac cones, which are well separated in $k$-space and with the Fermi level lying below the Lifshitz transition.

The first detailed magneto-transport study of Cd$_3$As$_2$ dates back to the 1960s~\cite{RosenmanPL66,ArmitagePLA68,RosenmanJPCS69}. The geometry of the Fermi surface was for the first time addressed by Shubnikov-de Haas (SdH) measurements.  Rosenman~\cite{RosenmanPL66} explored such quantum oscillations on a series of $n$-doped samples, concluding that the Fermi surface is a simple ellipsoid symmetric around the $c$-axis, and inferred a low anisotropy factor of 1.2. These very first papers also show nearly conical shape of the conduction band (Fig.~\ref{Rosenman}). A decade later, Zdanowicz~\emph{et al.}~\cite{ZdanowiczTSF79} studied SdH oscillations in thin films and single crystals of  Cd$_3$As$_2$, confirming the ellipsoidal geometry of the Fermi surface. In addition, they reported a striking linear magnetoresistance (MR).

\begin{figure*}
	\includegraphics[width=0.9\textwidth,trim={0cm 0cm 0cm 0cm}]{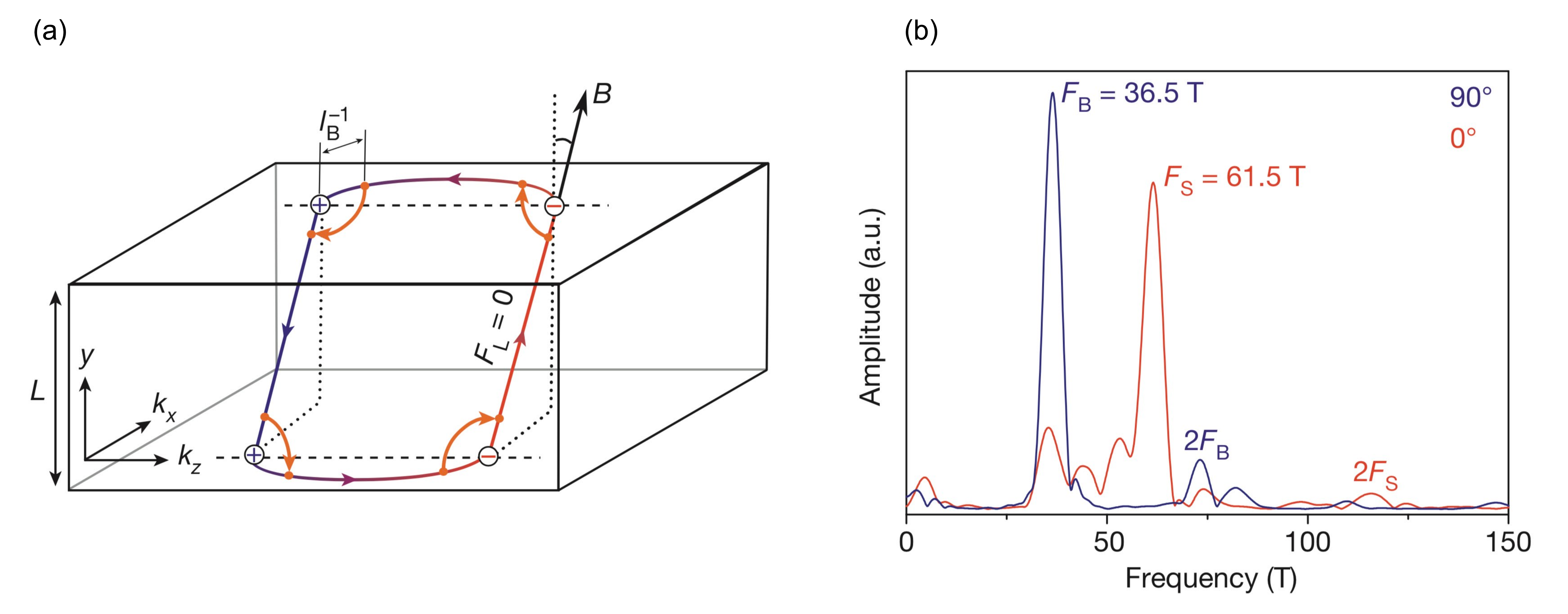}
	\caption{(a) Weyl orbit in a thin slab of thickness $L$ in a magnetic field $B$. The orbit involves the Fermi-arc surface states connecting the Weyl nodes of opposite chirality, and the bulk states of fixed chirality.
	(b) The Fourier transform of magnetoresistance at $T=2$~K measured on a thin Cd$_3$As$_2$ sample ($L=150$ nm) which was cut from a single crystal by focused ion beam. The field was oriented  parallel (90$^\circ$) and perpendicular (0$^\circ$) to the surface. Only one frequency is observed for parallel field, and two frequencies for perpendicular field.
	 Adapted by permission from Springer Nature: Nature~\cite{MollNature16}, copyright (2016).
	\label{Moll}}
\end{figure*}

In a new bout of activity, several groups confirmed that the Fermi surface consists of a simple, nearly spherical ellipsoid, with an almost isotropic Fermi velocity \cite{LiangNatureMater14}. Such a simple Fermi surface was questioned by Zhao~\emph{et al.}~\cite{ZhaoPRX15}. They found that, for particular directions of the magnetic field with respect to the main crystal axes, the magnetoresistance (MR) shows two oscillation periods which mutually differ by 10-25\%, pointing to a dumbbell-shaped Fermi surface. This was interpreted to originate from two nested Fermi ellipsoids arising from two separated Dirac cones, where $E_F$ is placed just above the Lifshitz transition. Another picture was proposed by Narayanan~\emph{et al.}~\cite{NarayananPRL15}, who found single-frequency SdH oscillations, and concluded two nearly isotropic Dirac-like Fermi surfaces. In subsequent studies, Desrat~\emph{et al.} followed the SdH oscillations as a function of the field orientation, and always found two weakly separated frequencies (5-10\%). Their results were interpreted within the picture of two ellipsoids that are separated in $k$-space due to the possible absence of inversion symmetry~\cite{DesratPRB18}.
Clear beating patterns, indicating a multiple frequency in SdH oscillations, were also reported in Nernst measurements~\cite{LiangPRL17}, and magnetoresistance~\cite{GuoSR16} on single crystals of Cd$_3$As$_2$. Such beating patterns were attributed to the lifting of spin degeneracy due to inversion symmetry breaking either by an intense magnetic field~\cite{XiangPRL15,LiangPRL17}, or by Cd-antisite defects~\cite{GuoSR16}, which may turn a Dirac node into two Weyl nodes.

Several transport studies reported on strikingly linear nonsaturating MR in Cd$_3$As$_2$~\cite{NarayananPRL15,FengPRB15,LiangNatureMater14} being more pronounced in samples with lower mobilities.
Since linear MR is observed at magnetic fields far below the quantum limit, the standard Abrikosov's theory~\cite{AbrikosovJPAMG03} -- referring to the transport in the lowest Landau level -- cannot be applied, and a different explanation was needed. Liang~\emph{et al.} \cite{LiangNatureMater14} therefore suggest there is a mechanism that protects from backscattering in zero field. This protection is then rapidly removed in field, leading to a very large magnetoresistance. They propose an unconventional mechanism, caused by the Fermi surface splitting into two Weyl
pockets in an applied magnetic field.
Similarly, Feng~\emph{et al.}~\cite{FengPRB15} assigned the large non-saturating MR to a lifting of the protection against backscattering, caused by a field-induced change in the Fermi surface. The authors judged that linear MR cannot be due to disorder, as the Cd$_3$As$_2$ samples are high-quality single crystals. They instead conclude that it is due to the Dirac node splitting into two Weyl nodes. In contrast, Narayanan~\emph{et al.}~\cite{NarayananPRL15} find that the Fermi surface does not significantly change up to 65 T, except for Zeeman splitting caused by a large $g$-factor. Through comparing quantum and transport relaxation times, they conclude that transport in Cd$_3$As$_2$ is dominated by small-angle scattering, which they trace back to electrons scattered on arsenic vacancies, and that the linear MR is linked to mobility fluctuations.

\begin{figure*}
	\includegraphics[width=0.9\textwidth,trim={0cm 0cm 0cm 0cm}]{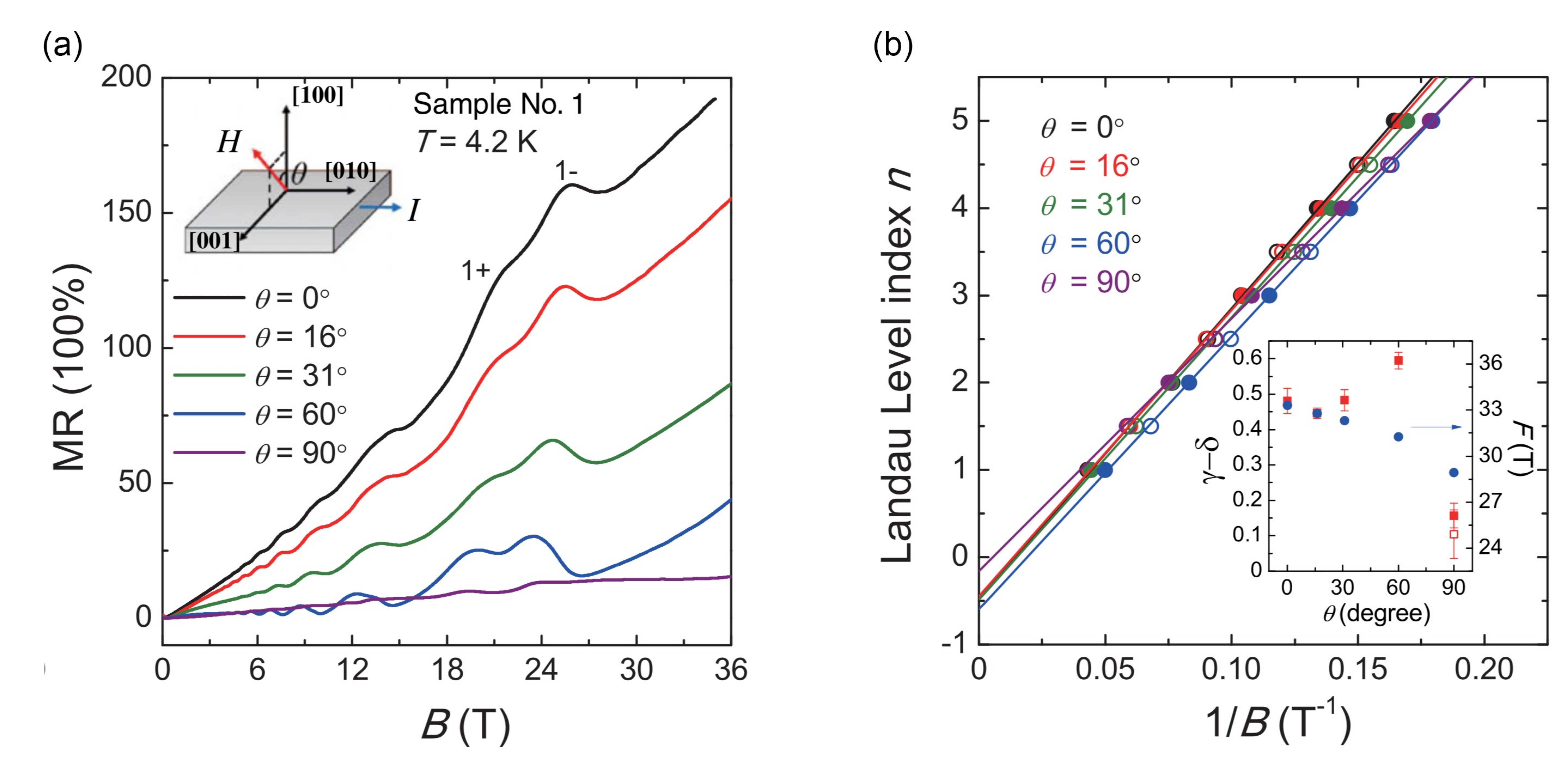}
	\caption{(a) Angular dependence of the transverse magnetoresistance with magnetic field $B$ rotated in the (010) plane. The tilt angle $\theta$ is the angle between $B$ and the [100] direction.
(b) Landau index plots $n$ vs $1/B$ at different $\theta$'s. (Inset) Angular dependence of the oscillation frequency $F$ and the total phase $\gamma - \delta$, extracted from the linear fitting from the first Landau level to the fifth Landau level.
Adapted with permission from~\cite{XiangPRL15}, copyright (2015) by the American Physical Society. \label{Xiang}}
\end{figure*}

The Potter~\emph{et al.} theory~\cite{PotterNatureComm14} predicts the existence of specific closed cyclotron orbits in a Dirac or Weyl semimetal~(Fig.\ref{Moll}a). These orbits are composed of two Fermi arcs located on opposite surfaces of the sample, which are then interconnected via zero-mode Landau levels. Moll~\emph{et al.}~\cite{MollNature16} report on the SdH oscillations measured in mesoscopic devices. These were prepared using the focused ion beam technique, allowing Cd$_3$As$_2$ crystals to be cut into sub-micron-thick platelets. They report two series of oscillations, and via specific angle dependence, they associate them with surface-related and bulk-related orbits (Fig.~\ref{Moll}b).

The chiral anomaly is yet another theoretically expected signature of the field-induced splitting of a Dirac point into a pair of Weyl nodes. The parallel application of an electric and magnetic field is predicted to transfer electrons between nodes with opposite chiralities. Such a transfer should be associated with lowering resistivity (negative MR).
For Cd$_3$As$_2$, there are indeed several reports of a negative MR and a suppression of thermopower for particular magnetic-field directions \cite{LiNatureComm15,JiaNatureComm16,LiNatureComm16}. Typically, such behavior was observed in microdevices (platelets or ribbons). It should be noted that a negative MR can also emerge in a topologically trivial case due to so-called current jetting~\cite{dosReisNJP16,LiangPRX18}. This is a simple consequence of a high transport anisotropy when a magnetic field is applied. In such a case, the current is jetting through a very narrow part of the sample. The measured MR then strongly depends on the distance of the voltage contacts from the current path.

The analysis of the quantum oscillation phase represents a unique way to identify the nature of probed charge carriers.
For conventional Schr\"{o}dinger electrons, one expects so-called Berry phase $\beta=0$, whereas massless Dirac electrons should give rise to $\beta = \pi$. Indeed, several reports based
on SdH oscillations in Cd$_3$As$_2$ indicate that $\beta \sim \pi$~\cite{ZdanowiczTSF79,HePRL14,DesratJPCS15,PariariPRB15}. A weak deviation from the ideal non-trivial Berry
phase, $\beta=(0.8-0.9)\pi$ was reported by Narayanan~\cite{NarayananPRL15}, who also discussed the influence of Zeeman splitting on the phase
of quantum oscillations~\cite{MikitikPRB03}. Notably, the electron $g$-factor in Cd$_3$As$_2$ is relatively large and anisotropic, and it implies Zeeman splitting comparable to the cyclotron energy~\cite{BlomSSC80,Blom80}.

Contrasting results were obtained by Xiang~\emph{et al.}~\cite{XiangPRL15},
who found the non-trivial Berry phase only when the field is applied along the tetragonal $c$-axis of Cd$_3$As$_2$. When the magnetic field is rotated to be parallel with the $a$ or $b$ axis, the measured Berry phase becomes nearly trivial (see Fig.~\ref{Xiang} and discussion in Ref.~\cite{HeCPB16}). This result is interpreted in terms of 3D Dirac-phase symmetry-breaking effects, when the magnetic field is tilted away from the $c$-axis. A certain angle dependence of the Berry
phase has also been reported by Zhao~\emph{et al.}~\cite{ZhaoPRX15}, reporting the Berry phase in between 0 and $\pi$. Another study by Cheng~\emph{et al.}~\cite{ChengNJP16}
was dedicated to the thickness-dependence of the Berry phase in MBE-grown thin films of Cd$_3$As$_2$, implying a non-trivial to trivial transition of the Berry phase with changing the layer thickness.

\begin{figure}[b]
	\includegraphics[width=0.45\textwidth,trim={0cm 0cm 0cm 0cm}]{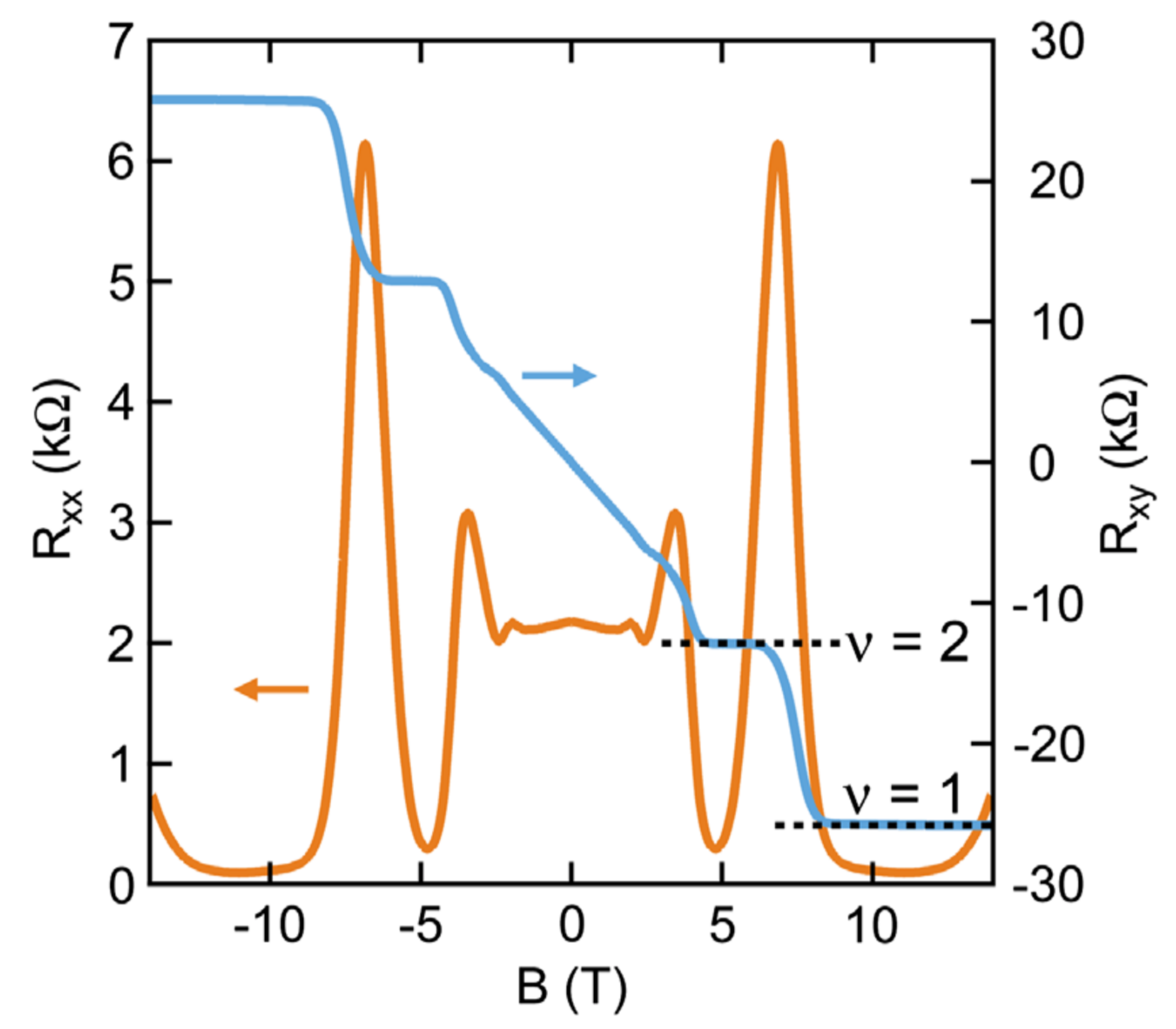}
	\caption{The quantum Hall effect in a 20-nm-thick epitaxial Cd$_3$As$_2$ film grown by molecular beam epitaxy, showing the Hall ($R_{xy}$) and longitudinal ($R_{xx}$) resistances as a function of magnetic field measured at 1~K. Reprinted with permission from~\cite{SchumannPRL18}, copyright (2018) by the American Physical Society.\label{Schumann}}
\end{figure}

The partly contradicting results in determining the Berry phase illustrate that even though the phase of quantum oscillation in principle identifies the nature unambiguously, the practical analysis of this phase is straightforward only in well-defined system such as graphene~\cite{NovoselovNature05}. In more complex materials such as Cd$_3$As$_2$, where quantum oscillations are superimposed on the much stronger effect of linear magneto-resistance~\cite{LiangNatureMater14}, the precise and reliable determination of the phase may represent a more challenging task. This may be illustrated with the example of bulk graphite, which is another high-mobility system with a strong and approximately linear magneto-resistance. There, the Dirac-like or normal massive nature of hole-type carriers has been a subject of intensive discussion in literature~\cite{LukyanchukPRL04,LukyanchukPRL06,SchneiderPRL09,LukyanchukPRL10,SchneiderPRL10}.

More recently, the very first experiments showing the quantum Hall effect (QHE) in thin films of Cd$_3$As$_2$ appeared. Zhang~\emph{et al.} measured SdH oscillations on a series of Cd$_3$As$_2$ nanoplates~\cite{ZhangNatureComm17}. They report multiple cyclotron orbits, distinguishing both 3D and 2D Fermi surfaces. They also
observe a quantized Hall effect (QHE), which they attribute to the surface states of Cd$_3$As$_2$, linked to the Weyl orbits. Schumann~\emph{et al.}~\cite{SchumannPRL18} studied MBE-grown, 20~nm thick films of Cd$_3$As$_2$, and observed the QHE at low temperatures (Fig.~\ref{Schumann}). Similarly, they attribute the QHE to surface states, and conclude that the bulk states are weakly gapped at low temperatures. In continuation of this work, Galletti~\emph{et al.}~\cite{GallettiPRB18} tuned the carrier concentration in thin films across the charge neutrality point using a gate voltage and concluded that the observed magneto-transport response is in line with expectations for a 2D electron gas of massless Dirac electrons.

The planar Hall effect has also been reported in Cd$_3$As$_2$, an effect in which a longitudinal current and an in-plane magnetic field give rise to a transverse current or voltage.
Li~\emph{et al.} investigated microribbons of Cd$_3$As$_2$, and found an anisotropic MR and planar Hall effect, which they attribute to the physics of Berry curvature~\cite{LiPRB18}.
Guo~\emph{et al.} also uncovered unusually large transverse Hall currents in needle-like single crystals of Cd$_3$As$_2$~\cite{GuoSR16}.
Wu~\emph{et al.} carried out similar measurements on Cd$_3$As$_2$ nanoplatelets, finding a large negative longitudinal MR, and a planar Hall effect with non-zero transverse voltage when the magnetic field is tilted away from the electric field. These observations are interpreted as transport evidence for the chiral anomaly~\cite{WuPRB18}.

\section{Summary}

Cadmium arsenide is a prominent member of the topological materials class, widely explored using both theoretical and experimental methods. At present, there is no doubt that Cd$_3$As$_2$ hosts well-defined 3D massless charge carriers. These are often associated with the symmetry-protected 3D Dirac phase, but may also be interpreted using alternative approaches developed in the past. This calls for further investigations of Cd$_3$As$_2$, preferably using a combination of different experimental techniques. Such investigation may resolve the currently existing uncertainties about the electronic band structure, which slow down the overall progress in physics of Cd$_3$As$_2$, and may contribute to our understanding of rich phenomena associated with this appealing material.

\vspace{2mm}
\subsection*{Acknowledgements}
The authors acknowledge discussions with F.~Bechstedt, S.~Borisenko, C.~C.~Homes, S.~Jeon, N.~Miller, B.~A.~Piot, A.~V.~Pronin, O.~Pulci, A.~Soluyanov, S.~Stemmer, Z. Wang, H. Weng and F. Xiu. This work was supported by ANR DIRAC3D projects and MoST-CNRS exchange programme (DIRAC3D). A.~A. acknowledges funding from The Ambizione Fellowship of the Swiss National Science Foundation. I.~C. acknowledges support from the postdoc mobility programme of the Suisse National Science Foundation. A.~A. acknowledges the support SNF through project PP00P2\_170544.


%

\end{document}